%% file: ForSin22Nov.tex
\documentclass[12pt]{article}
\newlength{\mytopmargin}
\newlength{\myleftmargin}
\newtheorem{theorem}{Theorem}[section]
\newtheorem{proposition}[theorem]{Proposition}
\newtheorem{cor}[theorem]{Corollary}
\newcommand{\Pf}{\ensuremath{\mathrm{Pf}}}
\newcommand{\sgn}{\ensuremath{\mathrm{sgn}}}
\newcommand{\Q}{\ensuremath{\mathbb{Q}}}
\newcommand{\transpose}{\ensuremath{\mathsf{T}}}

\usepackage{amsmath,amsfonts,amssymb}

\usepackage{graphicx}

\begin{document}
\title{A generalized plasma and interpolation between classical random matrix ensembles}
\author{Peter J. Forrester${}^\dagger$ and Christopher D. Sinclair${}^\#$}
\date{}
\maketitle

\noindent
\thanks{\small
${}^\dagger$ Department of Mathematics and Statistics,
The University of Melbourne,
Victoria 3010, Australia \\
email: { P.Forrester@ms.unimelb.edu.au} \\
${}^\#$ Department of Mathematics,  University of Oregon, 
Eugene OR 97403 \\
email: {csinclai@uoregon.edu}
}

\begin{abstract}
\noindent
The eigenvalue probability density functions of the classical random matrix ensembles have a well known analogy with the one component log-gas at the special couplings $\beta = 1,2$ and 4. It has been known for some time that there is an exactly solvable two-component log-potential plasma which interpolates between the $\beta =1$ and 4 circular ensemble, and an exactly solvable two-component generalized plasma which interpolates between $\beta = 2$ and 4 circular ensemble. We extend known exact results relating to the latter --- for the free energy and one and two-point correlations --- by giving the general $(k_1+k_2)$-point correlation function in a Pfaffian form. Crucial to our working is an identity which expresses the Vandermonde determinant in terms of a Pfaffian. The exact evaluation of the general correlation is used to exhibit a perfect screening sum rule.
\end{abstract}

\section{Introduction}
\subsection{The circular ensembles of Dyson}
In classical random matrix theory (see e.g.~\cite{Fo10}), the probability density function (PDF)
\begin{equation}\label{1}
{1 \over C_{N,\beta}} \prod_{1 \le j < k \le N} |e^{i \theta_k} - e^{i \theta_j}|^\beta
\end{equation}
for $\beta = 1,2$ and 4 plays a distinguishing role. Thus (\ref{1}) for these values of $\beta$
is the eigenvalue PDF for Dyson's circular ensembles of unitary random matrices
$\{ U_N U_N^T \}$, $\{ U_N\}$, $\{U_{2N} U_{2N}^D\}$ respectively. Here $U_N \in U(N)$,
chosen with Haar measure and $U^D = Z_{2N} U^T Z_{2N}^{-1}$, where
$Z_{2N} = {I}_N \otimes \begin{bmatrix} 0 & -1 \\ 1 & 0 \end{bmatrix}$
denotes the quaternion dual. The spectrum of matrices $U_{2N} U_{2N}^D$ is doubly degenerate,
which explains why (\ref{1}) still involves $N$ eigenvalues for $\beta = 4$.

Dyson \cite{Dy62} introduced the circular ensembles as alternatives to the Gaussian ensembles
of Wigner. Unlike the latter, they are uniquely determined by an invariance property: their measure
is unchanged under left and right conjugation by real orthogonal, unitary, and symplectic unitary
matrices respectively. Significantly, it was found by explicit calculation that in the bulk scaling limit the two-point correlation functions for the two classes of ensembles are identical. Extrapolating this to the conjecture that for a given $\beta$, all statistical properties of the bulk states of the circular and Gaussian ensembles are identical was important at the time, as it allowed calculations to be carried out on the computationally simpler circular ensembles, with the aim of answering questions relating to Gaussian Hermitian matrices.

A celebrated example is the statement that \cite[eq.~18.160]{Fo10}
\begin{equation}\label{2}
p_4^{\rm bulk}(k;s) = 2 p_1^{\rm bulk}(2k+1;2s)
\end{equation}
where $p_\beta^{\rm bulk}(k;s)$ denotes the PDF that a gap of size $s$ between two eigenvalues
contains $k$ eigenvalues in the bulk state at coupling $\beta$. This was deduced by deriving for the 
circular ensembles that \cite{MD63}, \cite[eq.~(8.157)]{Fo10}
\begin{equation}
p_{N,4}(n;(-\theta,\theta)) = p_{2N,1}(2n+1;(-\theta,\theta)).
\end{equation}
Here the notation $p_{N,\beta}(k;-(\theta,\theta))$ is analogous to $p_\beta^{\rm bulk}(k;s)$ but with
the extra subscript $N$ referring to the total number of eigenvalues.

The proof of the conjecture that for $\beta=1,2$ and 4 the bulk states of the circular and Gaussian ensembles are the same became possible after another landmark paper of Dyson \cite{Dy70}.
In this work a formalism was given which allowed not just for the calculation of the 2-point
correlation functions, but the general $k$-point correlations $\rho_{(k)}$. This formalism was applied to the circular ensembles. In the bulk scaling limit, particle density equal to unity,  it gave
\begin{equation}\label{3}
\rho_{(k)}(x_1,\dots,x_k) = \det \Big [ {\sin \pi (x_j - x_l) \over \pi (x_j - x_l)}
\Big ]_{j,l=1,\dots,k}
\end{equation}
for $\beta = 2$, and
\begin{equation}\label{4}
\rho_{(k)}(x_1,\dots,x_k) =
 {\rm qdet} \left[
\begin{array}{ll}
\displaystyle{{\sin 2\pi(x_j-x_l)
\over 2 \pi(x_j-x_l)}} & \displaystyle{
{1 \over 2 \pi }
\int_0^{2 \pi (x_j-x_l)} {\sin  x \over  x} \, dx} \\[.4cm]
\displaystyle{{\partial \over
\partial x_j} {\sin 2\pi\rho(x_j-x_l) \over 2 \pi
(x_j-x_l)}} & \displaystyle{{\sin 2\pi(x_j-x_l)
\over 2 \pi(x_j-x_l)}}
\end{array} \right]_{j,l=1,\dots,k}
\end{equation}
for $\beta = 4$ (a formula analogous to (\ref{4}) was also given for $\beta = 1$, but as it plays no
role in our subsequent discussion we refrain from writing it down). The $2k \times 2k$ matrix,
$S$ say, in (\ref{4}), has the property of being self dual: $S^D := Z_{2k} S^T S_{2k}^{-1} = S$.
It follows that $S Z_{2 k}^{-1}$ is antisymmetric so its Pfaffian is well defined. The quaternion
determinant in (\ref{4}) is then given in terms of a Pfaffian according to ${\rm qdet} \, S =
{\rm Pf} (S Z_{2k}^{-1})$. The method of \cite{Dy70} was subsequently applied to the Gaussian ensembles \cite{Me71}, allowing in particular the verification that the bulk statistical properties coincide with that of the circular ensemble.

\subsection{From the circular ensembles to two-component log-gases and trial wave functions}
As emphasized by Dyson \cite{Dy62}, for general $\beta > 0$ (\ref{1}) has the interpretation as the Boltzmann factor for a classical statistical mechanical system with $N$ particles on a circle interacting via the pair potential $- \log |e^{ i \theta} - e^{i \phi}|$ (one-component log-gas). Sutherland
\cite{Su71} showed that (\ref{1}) is the absolute value squared of the ground state wave function for the quantum many body system specified by the Hamiltonian
\begin{equation}\label{4.1}
- \sum_{j=1}^N {\partial^2 \over \partial \theta_j^2} +
{\beta \over 4} \Big ( {\beta \over 2} - 1 \Big )
\sum_{1 \le j < k \le N} {1 \over \sin^2 \pi (\theta_k - \theta_j)/2}.
\end{equation}
These alternative interpretations suggest two-component generalizations of (\ref{1}) which
interpolate between the classical couplings $\beta = 1,2$ and 4.

One such generalization is \cite{Fo84a}
\begin{equation}\label{4.2}
\prod_{1 \le j < k \le N_1} |e^{i \theta_k} - e^{i \theta_j} |
\prod_{1 \le \alpha < \beta \le N_2} |e^{i \phi_\alpha} - e^{i \phi_\beta} |^4
\prod_{j=1}^{N_1} \prod_{\alpha=1}^{N_2} |e^{i \theta_j} - e^{i \phi_\alpha} |^2.
\end{equation}
It is the Boltzmann factor for a two-component log-potential plasma system on the circle with $N_1$ particles of charge
$+1$, and $N_2$ particles of charge $+2$. It is also the ground state wave function for a
two-component generalization of (\ref{4.1}) \cite{KO82}. For $N_2 = 0$ ($N_1=0$)
(\ref{4.2}) reduces to (\ref{1}) with $\beta = 1$ ($\beta = 4$). Moreover, the general 
$(k_1,k_2)$-point correlation function can be computed in a quaternion determinant form
\cite[eq.~(6.168)]{Fo10} which interpolates between the quaternion determinant forms for
$\beta =1$ and $\beta = 4$. And in a very recent development the Gaussian analogue of
(\ref{4.2}) has shown itself to be exactly solvable \cite{RS10}.

A second such generalization is \cite{FJ84}
\begin{equation}\label{5.1}
{1 \over C(N_1,N_2)} \prod_{1 \le j < k \le N_1} |e^{i \theta_k} - e^{i \theta_j} |^2
\prod_{1 \le \alpha < \beta \le N_2} |e^{i \phi_\alpha} - e^{i \phi_\beta} |^4
\prod_{j=1}^{N_1} \prod_{\alpha=1}^{N_2}  |e^{i \theta_j} - e^{i \phi_\alpha} |^2.
\end{equation}
This describes a Boltzmann factor of a so-called generalized plasma: the couplings (exponents
on the products of differences) $g_{RR}$, $g_{GG}$, $g_{RG}$ for the two species (R)oman
and (G)reek no longer satisfy $g_{RR} g_{GG} = g_{RG}^2$ as they do in (\ref{4.2}).
With the complex exponentials replaced by general complex coordinates, this is of interest as
a trial wave function in the anomolous quantum Hall effect \cite{Ha83}, and without such replacement as the exact wave function for a certain supersymmetric $tJ$ system in one-dimension
\cite{AKS04,KK09}. For $N_2 = 0$ $(N_1 = 0)$, (\ref{5.1}) reduces to (\ref{1}) with $\beta = 2$ ($\beta = 4$).
However, unlike the situation with the two-component system (\ref{4.2}), there is no existing formula in the literature for the general $(k_1,k_2)$-point correlation function, nor are there any exact
calculations carried out for the Gaussian analogue of (\ref{5.1}).

Two exact solvability features of (\ref{5.1}) are in the existing literature. One is the exact evaluation of the normalization,
\begin{equation}\label{6.1}
C(N_1,N_2) = (2 \pi)^{N_1 + N_2} {(2N_2 + N_1)! (N_1/2)! N_2 ! \over
2^{N_2} (N_2 + N_1/2)!},
\end{equation}
and the other is an exact evaluation of the two-point correlations. The latter, in the bulk scaling
limit with density of Roman species $\rho_R$ and Greek species $\rho_G$, read \cite{FJ84}
\begin{eqnarray}
&&
\rho_{RR}(x,0) = \rho_R^2 - {\sin^2 \pi \rho_R x \over (\pi x)^2}  \nonumber \\
&& \qquad \qquad -
{\rho_G  \rho_R} \int_0^1 dt \int_0^1 ds \, {1 - 2s \over 2 \rho_G t/\rho_R + 1}
\Big ( e^{2 \pi i \rho_R x (s - 1) - 2 \pi i \rho_G x t} -
e^{2 \pi i \rho_R x s + 2 \pi i \rho_G xt} \Big ) \label{10}\\
&&\rho_{GG}(x,0) = \rho_G^2 + \rho_G^2 \int_0^1dt \int_0^1 ds \, \bigg ( -
{(t + s + \rho_R/\rho_G)^2 \over (2t + \rho_R/\rho_G) (2 s + \rho_R/\rho_G)}
e^{2 \pi i \rho_G x (t - s)} \nonumber \label{11} \\
&& \qquad \qquad  + {(t - s)^2 \over (2t + \rho_R/\rho_G)(2s+\rho_R/\rho_G)}
\cos \Big ( 2 \pi \rho_G x (t +s + \rho_R/\rho_G) \Big ) \bigg )
\\
&& \rho_{RG}(x,0) = \rho_R \rho_G - 2  \rho_R \rho_G  \int_0^1 dt \int_0^1 ds \,
{\rho_G t + \rho_R s \over 2 \rho_G t + \rho_R}
\cos\Big (2 \pi \rho_G xt - 2 \pi \rho_R x (s - 1) \Big ), \label{12}
\end{eqnarray}
which can of course be equally as well written as products of one-dimensional integrals.
In the case $(\rho_G,\rho_R) = (1,0)$ the first of these reduces to (\ref{3}) with $k=2$, while in
the case $(\rho_G, \rho_R) = (0,1)$ the second reduces to (\ref{4}) with $k=2$.

\subsection{Our aim}
In light of the above discussions, two challenges present themselves.
\begin{enumerate}
\item To compute the $(k_1,k_2)$-point correlation function corresponding to the
generalized two-component plasma (\ref{5.1}) in a form which relates to (\ref{3}) and (\ref{4}).
\item To similarly compute the $(k_1,k_2)$-point correlation function for the Gaussian analogue
of (\ref{5.1}).
\end{enumerate}
We will see how application of a Pfaffian structure only very recently introduced into random
matrix theory \cite{Si10} allows us to solve the first of these problems. 
In the Appendix we generalize this Pfaffian formula to a form suitable for application to the second problem, but we leave the required subsequent working to a later work.

The previously computed two-point correlations (\ref{10})--(\ref{12}) hold in the bulk scaling limit. We obtain the corresponding $(k_1,k_2)$-point correlation functions by taking the bulk
scaling  limit of our finite system results. In the work \cite{FJ84}
some special screening properties of the bulk scaled truncated two-point functions,
\begin{equation}\label{13}
\int_{-\infty}^\infty \rho_{RR}^T(x,0) \, dx = - \rho_R, \quad
\int_{-\infty}^\infty \rho_{RG}^T(x,0) \, dx = 0, \quad
\int_{-\infty}^\infty \rho_{GG}^T(x,0) \, dx = -\rho_G
\end{equation}
were noted. Knowledge of the $(k_1,k_2)$-point correlation function allows these sum rules to be generalized so as to relate to the truncated form of the general correlation functions.
Furthermore, we give a discussion relating to the correlations in the large $\rho_R$ or
large $\rho_G$ limit.

\section{Correlations for the circular generalized plasma}
\subsection{Generalized partition function}
The method used in \cite{FJ84} to deduce the exact two-point correlations (\ref{10})--(\ref{12})
was to write (\ref{6.1}) as the product of a Vandermonde and confluent Vandermonde determinant,
by noting that
\begin{align}
\prod_{1 \le j < k \le N_1} (z_k - z_j) & = \det [ z_j^{k-1}]_{j,k=1,\dots,N_1}
\label{V1} \\
\prod_{1 \le j < k \le N_1+N_2} (z_k - z_j)^{q_j q_k} &=
\det \begin{bmatrix} [z_j^{k-1}]_{j=1,\dots,N_1 \atop k=1,\dots,N_1+ 2N_2} \\
\begin{bmatrix} z_j^{k-1} \\ (k-1) z_j^{k-2}  \end{bmatrix}_{j=N_1+1,\dots,N_1+N_2 \atop 
k=1,\dots,N_1 + 2N_2} \end{bmatrix} \label{V2}
\end{align}
where $q_j = 1$ $(j=1,\dots,N_1)$, $q_j=2$ ($j=N_2+1,\dots,N_1+N_2$),
respectively.

The most crucial strategic step of the present study is to replace (\ref{V1}) by the Pfaffian identity
\cite{IO06}
\begin{equation}\label{V3}
\prod_{1 \le j < k \le N_1} (z_k - z_j) = {\rm Pf} \, \Big [ {(z_k^{N_1/2} - z_j^{N_1/2})^2 \over
z_k - z_j} \Big ]_{j,k=1,\dots,N_1}.
\end{equation}
We give a new proof of this identity, and a generalization which is of potential future use in
random matrix theory in the Appendix.
This structure was introduced into random matrix theory in the recent work of one of us
\cite{Si10}. It allows us to express the generalized partition function
\begin{eqnarray}
&&  Z_{N_1,N_2}[u,v] := {1 \over C(N_1,N_2)} \int_0^{2 \pi} d \theta_1 \, u(\theta_1)
\cdots \int_0^{2 \pi} d \theta_{N_1} \, u(\theta_{N_1})  \int_0^{2 \pi} d \phi_1 \, v(\phi_1)
\cdots \int_0^{2 \pi} d \phi_{N_2} \, v(\phi_{N_2}) \, \nonumber \\
&& \quad \times 
 \prod_{1 \le j < k \le N_1} |e^{i \theta_k} - e^{i \theta_j} |^2
\prod_{1 \le \alpha < \beta \le N_2} |e^{i \phi_\alpha} - e^{i \phi_\beta} |^4
\prod_{j=1}^{N_1} \prod_{\alpha=1}^{N_2}  |e^{i \theta_j} - e^{i \phi_\alpha} |^2 \label{Z}
\end{eqnarray}
in terms of a Pfaffian.

\begin{proposition}
With $z:= e^{i \theta}$, and $\{p_{j-1}(z)\}_{j=1,2,\dots}$ an arbitrary set of monic polynomials,
$p_{j-1}(z)$ of degree $j-1$, let
\begin{eqnarray}
&& \displaystyle a_{j,k}[u] = \int_0^{2 \pi} d\theta_1 \, u(\theta_1)  \int_0^{2 \pi} d\theta_2 \, u(\theta_2) \,
z_1^{-N_2} z_2^{-N_2} {(z_{2}^{-N_1/2} - z_{1}^{-N_1/2})^2 \over
z_{2}^{-1} - z_{1}^{-1}}
 p_{j-1}(z_1) p_{k-1}(z_2) \label{A1} \\
&& b_{j,k}[v] = \int_0^{2 \pi}  v(\theta)  z^{-(N_1+2N_2 - 2)} \Big ( p_{j-1}(z) p_{k-1}'(z) -
p_{j-1}'(z) p_{k-1}(z) \Big ) \, d \theta. \label{B1}
\end{eqnarray}
In terms of this notation, and with $[\zeta^k] f(\zeta)$ denoting the coefficient
of $\zeta^k$ in the power series expansion of $f(\zeta)$, for $N_1$ even we have
\begin{equation}\label{Z7}
Z_{N_1,N_2}[u,v] =
{N_1 ! N_2! \over C(N_1,N_2)} [\zeta^{N_1/2}] {\rm Pf} \, [ \zeta a_{j,k}[u] + b_{j,k}[v] 
]_{j,k=1,\dots,N_1+2N_2}.
\end{equation}
\end{proposition}

\noindent
Proof. \quad Using the identity
$$
| e^{ i \theta_k} - e^{i \theta_j} | = -i e^{-i(\theta_j + \theta_k)} (e^{i \theta_k} - e^{i \theta_j})
$$
valid for $0 < \theta_j < \theta_k < 2 \pi$ the absolute values appearing in the integrand can be
eliminated, and we obtain
\begin{eqnarray}\label{Z1}
&& Z_{N_1,N_2}[u,v] := {1 \over C(N_1,N_2)} \int_0^{2 \pi} d \theta_1 \, u(\theta_1)
\cdots \int_0^{2 \pi} d \theta_{N_1} \, u(\theta_{N_1})  \int_0^{2 \pi} d \phi_1 \, v(\phi_1)
\cdots \int_0^{2 \pi} d \phi_{N_2} \, v(\phi_{N_2}) \, \nonumber \\
&& \quad \times 
\prod_{l=1}^{N_1} z_l^{-N_2} \prod_{\alpha = 1}^{N_2} z_{N_1+\alpha}^{-(N_1+2N_2 - 2)}
\prod_{1 \le j < k \le N_1} (z_k^{-1} - z_j^{-1})
\prod_{1 \le j < k \le N_1+N_2} (z_k - z_j)^{q_j q_k}. 
\end{eqnarray}

Consider the Pfaffian formula (\ref{V3}) with the replacements ($z_j \mapsto z_j^{-1}$).
It can rewritten
\begin{equation}\label{V6}
\prod_{1 \le j < k \le N_1} (z_k^{-1} - z_j^{-1}) =
{1 \over 2^{N_1/2} (N_1/2)! }
{\rm Asym} \, \prod_{j=1}^{N_1/2} {(z_{2j}^{-N_1/2} - z_{2j-1}^{-N_1/2})^2 \over
z_{2j}^{-1} - z_{2j-1}^{-1}}
\end{equation}
where Asym denotes the operation of antisymmetrization in $\{z_1,\dots,z_{N_1}\}$.
Substituting (\ref{V6}) in (\ref{Z1}) we see from the fact that the final product of differences
is antisymmetric in $\{z_1,\dots,z_{N_1}\}$ that all terms in the Asym operation contribute
equally. Thus
\begin{eqnarray}\label{Z2}
&& Z_{N_1,N_2}[u,v] := {1\over \tilde{C}(N_1,N_2)} \int_0^{2 \pi} d \theta_1 \, u(\theta_1)
\cdots \int_0^{2 \pi} d \theta_{N_1} \, u(\theta_{N_1})  \int_0^{2 \pi} d \phi_1 \, v(\phi_1)
\cdots \int_0^{2 \pi} d \phi_{N_2} \, v(\phi_{N_2}) \, \nonumber \\
&& \quad \times 
\prod_{l=1}^{N_1} z_l^{-N_2} \prod_{\alpha = 1}^{N_2} z_{N_1+\alpha}^{-(N_1+2N_2 - 2)}
\prod_{j=1}^{N_1/2} {(z_{2j}^{-N_1/2} - z_{2j-1}^{-N_1/2})^2 \over
z_{2j}^{-1} - z_{2j-1}^{-1}}
\prod_{1 \le j < k \le N_1+N_2} (z_k - z_j)^{q_j q_k},
\end{eqnarray}
where $\tilde{C}(N_1,N_2) = 2^{N_1/2} (N_1/2)! C(N_1,N_2)/N_1!$.

Next, we observe (as is standard) that for $\{p_j(x)\}_{j=0,1,2,\dots}$ a set of monic polynomials,
$p_j(x)$ of degree $j$, the determinant in (\ref{2}) can be rewitten to read
\begin{eqnarray*}
&& \det \begin{bmatrix} [p_{k-1}(z_j)]_{j=1,\dots,N_1 \atop k=1,\dots,N_1+ 2N_2} \\
\begin{bmatrix} p_{k-1}(z_j) \\ (k-1) p_{k-1}'(z_j)\end{bmatrix}_{j=N_1+1,\dots,N_1+N_2 \atop
k=1,\dots,N_1 + 2N_2}  \end{bmatrix} \\
&& \quad =
\sum_{Q \in S_{N_1+2N_2}} \varepsilon(Q)
\prod_{l=1}^{N_1} p_{Q(l) - 1}(z_j) \prod_{l=N_1+1}^{N_1+N_2}
p_{Q(l) - 1}(z_j) p_{Q(l)-1}'(z_j),
\end{eqnarray*}
where the equality follows from the definition of a determinant as a sum over permutations.
Substituting this for the final term in (\ref{Z2}), we see that the integrations factorize down to one
and two dimensional integrals, and we obtain
\begin{eqnarray}\label{Z7a}
Z_{N_1,N_2}[u,v] & = & {1 \over \tilde{C}(N_1,N_2)} 
\sum_{Q \in S_{N_1+2N_2}}  \varepsilon(Q) \prod_{j=1}^{N_1/2} a_{Q(2l-1),Q(2l)}[u]
\prod_{l=N_1/2+1}^{N_1/2 + N_2} \tilde{b}_{Q(2l-1),Q(2l)}[v]. \nonumber \\
& = & {N_1! N_2! \over C(N_1,N_2)} 
\sum_{Q \in S_{N_1+2N_2}} \! \! \!{}^*  \, \varepsilon(Q) \prod_{j=1}^{N_1/2} a_{P(2l-1),P(2l)}[u]
\prod_{l=N_1/2+1}^{N_1/2 + N_2} {b}_{Q(2l-1),Q(2l)}[v]. 
\end{eqnarray}
Here 
$$
\tilde{b}_{j,k}[v] = \int_0^{2 \pi} \, v(\theta)  z^{-(N_1+2N_2 - 2)} p_{j-1}(z) p_{k-1}'(z) \, d\theta
$$
and the asterisk denotes that the permutations are required to satisfy $P(2l) > P(2l-1)$
for each $l=1,\dots,N_1/2+N_2$ and $P(1) < P(3) < \cdots < P(N_1 + 2N_2 - 1)$ (imposing these conditions is how the second equality is obtained
from the first).
But by writing out the Pfaffian in (\ref{Z7}) as a sum over permutations according to the
general formula
$$
{\rm Pf} \, C = \sum_{Q \in S_{2N}} \!\!\!{}^* \, \varepsilon(Q) \prod_{l=1}^N c_{Q(2l-1),Q(2l)}
$$
for $C = [c_{jk}]_{j,k=1,\dots,2N}$ antisymmetric, we see it reduces to (\ref{Z7a}).
\hfill $\square$

\subsection{Skew orthogonal polynomials}
Let $R$ denote a particular permutation of $\{1,2,\dots,N_1+2N_2\}$. Let us
rearrange the rows and columns of the Pfaffian in (\ref{Z7}) to have the order of $R$,
and thus
\begin{equation}\label{W1}
Z_{N_1,N_2}[u,v] = {N_1! N_2! \over C(N_1,N_2)} [\zeta^{N_1/2}]
{\rm Pf} \, [ \zeta a_{R(j),R(k)}[u] + b_{R(j), R(k)}[v] ]_{j,k=1,\dots,N_1+2N_2}.
\end{equation}
We seek a choice of the permutation $R$, and a choice of the polynomials $\{p_j(z)\}$,
such that the skew inner product $\zeta a_{R(j),R(k)} + b_{R(j), R(k)}$ has the skew orthogonality
property
\begin{equation}\label{W2}
\zeta a_{R(n),R(m)}[1] + b_{R(n),R(m)}[1] = \left \{
\begin{array}{ll} r_{j-1}, & (n,m) = (2j-1,2j) \\
0, & {\rm otherwise},
\end{array} \right.
\end{equation}
for $R(n) < R(m)$.

\begin{proposition}\label{p2}
For $j=1,\dots,N_1/2$ set
\begin{equation}\label{W3a}
R(2j-1) = N_2 + j, \qquad R(2j) = N_2 + N_1 - j + 1 
\end{equation}
and for $j=N_1/2+1,\dots,N_1/2+N_2$ set
\begin{equation}\label{W3b}
R(2j-1) = j - N_1/2, \qquad R(2j) = 2N_2 + 3N_1/2 - j + 1. 
\end{equation}
Choose
\begin{equation}\label{W3c}
p_j(z) = z^j \qquad (j=0,1,\dots,N_1+2N_2 - 1).
\end{equation}
We then have that the skew orthogonality property (\ref{W2}) holds with
\begin{equation}\label{W3d}
r_{j-1} = \left \{
\begin{array}{ll} (2 \pi)^2 \zeta + 2 \pi  (N_1 + 1 - 2j), & j=1,\dots,N_1/2 \\
2 \pi (2N_2 + 2N_1 + 1 - 2j), & j=N_1/2+1,
\dots,N_1/2 + N_2.
\end{array} \right.
\end{equation}
\end{proposition}

\noindent
Proof. \quad With $w = e^{i \phi}$, let us write
\begin{equation}\label{W4a}
\Phi_j(z) := \int_0^{2 \pi} w^{-N_2} {(z^{-N_1/2} - w^{-N_1/2})^2 \over z^{-1} - w^{-1}}
p_j(w) \, d \phi.
\end{equation}
Use of the expansion
\begin{equation}\label{W4b}
{z^{-N_1/2} - w^{-N_1/2} \over z^{-1} - w^{-1}} = \sum_{p=0}^{N_1/2 - 1}
(z^{-1})^p (w^{-1})^{N_1/2 - 1 - p}
\end{equation}
shows that
\begin{equation}\label{W4c}
\Phi_j(z) = \left \{
\begin{array}{ll} 2 \pi \, {\rm sgn} (N_2 + N_1/2-1/2-j) z^{j+1-N_1 - N_2}, &
j=N_2,\dots,N_2 + N_1 - 1 \\
0, & {\rm otherwise}. \end{array} \right.
\end{equation}
But according to the definition (\ref{A1})
$$
a_{j,k}[1] = \int_0^{2 \pi} z^{-N_2 + k - 1} \Phi_{j-1}(z) \, d \theta.
$$
Making use of (\ref{W4c}) we therefore have
\begin{equation}\label{A2}
a_{j,k}[1] = \left \{
\begin{array}{ll} (2 \pi)^2 {\rm sgn} (N_2 + N_1/2 + 1/2 - j) \delta_{j+k-1-N_1-2N_2,0},
& j=N_2+1,\dots,N_2+N_1 \\
0, & {\rm otherwise}. \end{array} \right.
\end{equation}

On the other hand, from the definition (\ref{B1}) and (\ref{W3c}) we see that
\begin{equation}\label{B2}
b_{j,k}[1] = (k-j) \int_0^{2 \pi} z^{-(N_1 + 2N_2) + j + k - 1} \, d \theta =
2 \pi (k - j) \delta_{j+k-1-N_1-2 N_2,0}.
\end{equation}

In light of (\ref{A2}) and (\ref{B2}), we see that the skew inner product in (\ref{W2}) is nonzero
for $R(j) + R(k) = N_1 + 2N_2 + 1$ only. But (\ref{W3a}) and (\ref{W3b}) tell us that
for $j  < k$, this equation holds if and only if $j = 2j' - 1$ and $k=2j'$. In these latter cases
we have
\begin{eqnarray*}
&& \zeta a_{R(2j-1), R(2j)} + b_{R(2j-1), R(2j)}\nonumber \\
&& \qquad  =   \Big (
(2 \pi)^2  \zeta {\rm sgn} (N_2 + N_1/2 + 1/2 - R(2j-1)) + 2 \pi
(2 N_2 + N_1 + 1 - 2 R(2j-1)) \Big )
\end{eqnarray*}
for $j=1,\dots,N_1/2$ and
$$
\zeta a_{R(2j-1), R(2j)}[1] + b_{R(2j-1), R(2j)}[1] = 2 \pi (2 N_2 + N_1 + 1 - 2 R(2j-1)) 
$$
for $j=N_1/2 +1,\dots,N_1/2 + N_2$,
which establishes the result. \hfill $\square$

\smallskip
According to (\ref{W1})
$$
Z_{N_1,N_2}[1,1] = {N_1! N_2! \over C(N_1,N_2)} [\zeta^{N_1/2}] \prod_{j=1}^{N_2/2+N_2} r_{j-1}.
$$
Substituting (\ref{W3d}) and simplifying we see that for $Z_{N_1,N_2}[1,1]=1$ the 
normalization must be given by (\ref{6.1}).

\subsection{Pfaffian form for the correlations}
Consider a two-component system of (R)oman and (G)reek particles. Let the one-body
potentials $u$ and $v$ be associated to each species respectively, and suppose the
corresponding generalized partition function has the form
\begin{equation}\label{Ma0}
\tilde{Z}_N[u,v] = {\rm Pf} [ \tilde{a}_{j,k}[u] + \tilde{b}_{j,k}[v] ]_{j,k=1,\dots,N}.
\end{equation}
Here $N$ is assumed even and $\tilde{a}_{j,k}[u]$, $\tilde{b}_{j,k}[v] $ are antisymmetric in
$j,k$.

Generally for a two-component system the $(k_1,k_2)$-point correlation function can be
computed from the generalized partition function by functional differentiation according to
the formula
\begin{eqnarray} \label{Ma0a}
&&
\rho_{(k_1,k_2)}(x_1,\dots,x_{k_1};y_1,\dots,y_{k_2}) \nonumber \\
&& \qquad =
{1 \over \tilde{Z}_N[1,1]} {\delta^{k_1 + k_2} \over
\delta u(x_1) \cdots \delta u(x_{k_1}) \delta v(y_1) \cdots \delta v(y_{k_2})}
\tilde{Z}_N[u,v] \Big |_{u=v=1}.
\end{eqnarray}
Our first point is that with some additional structure, the generalized partition function
(\ref{Ma0}) substituted into (\ref{Ma0a}) yields a Pfaffian of size $2(k_1+k_2)$.

Thus we suppose that 
\begin{eqnarray} \label{Ma1}
&& \tilde{a}_{j,k}[u] = 2 \int_I u(x) \psi_j^{(R)} \epsilon_R(u \psi_k^{(R)} )[x] \, dx \nonumber \\
&& \tilde{b}_{j,k}[v] = 2 \int_{\mathcal D} v(y) \Big ( \psi_j^{(G)}(y) 
\epsilon_G(\psi_k^{(G)})[y] - \psi_k^{(G)}(y) 
\epsilon_G(\psi_j^{(G)})[y]  \Big ) dy,
\end{eqnarray}
where, with $s \in \{R,G\}$, the $\epsilon_s$ are linear operators.
We see that (\ref{Z7}) is of the form (\ref{Ma1}) with $I = {\mathcal D} = [0,2\pi]$ and
\begin{eqnarray}
&&\psi_j^{(R)}(x) = z^{-N_2} p_{R(j+1)-1}(z), \qquad z = e^{2 \pi i x} \label{38} \\
&& \epsilon_R(f)[x] = {\zeta \over 2} \int_0^{2 \pi} {(z_2^{-N_1/2} - z^{-N_1/2})^2 \over
z_2^{-1} - z^{-1}} f(z_2) \, dz_2 \label{39} \\
&&\psi_j^{(G)}(x) = z^{-(N_1+2N_2 - 2)/2} p_{R(j+1) - 1}(z) \label{40}  \\
&& \epsilon_G(f)[y] = {1 \over 2} {d \over dz} f(z) \Big |_{z = e^{2 \pi i y}}. \label{41}
\end{eqnarray}
Furthermore we know from Proposition \ref{p2} that there is a choice of $\{p_j(x)\}$
which skew diagonalizes the inner product. Let us assume that the $\psi_j^{(s)}$ are
chosen so that the inner product implied by (\ref{Ma0}) with $u=v=1$ is similarly
skew-diagonalized. In this setting the explicit form of the corresponding $(k_1,k_2)$-point
correlation function (\ref{Ma0a}) has been computed by Mays \cite{Ma10} (see also
\cite{BS07}).

To present the result, let
\begin{eqnarray} \label{Ma4}
&& S_{s_1,s_2}(\mu,\eta) = 2 \sum_{j=0}^{N/2 - 1} {1 \over r_j} \Big (
\psi_{2j}^{(s_1)}(\mu) \epsilon_{s_2}(\psi_{2j+1}^{(s_2)})[\eta] -
\psi_{2j+1}^{(s_1)}(\mu) \epsilon_{s_2}(\psi_{2j}^{(s_2)})[\eta]  \Big ) \nonumber \\
&& D_{s_1,s_2}(\mu,\eta) = 2 \sum_{j=0}^{N/2 - 1} {1 \over r_j} \Big (
\psi_{2j}^{(s_1)}(\mu) \psi_{2j+1}^{(s_2)}(\eta)-
\psi_{2j+1}^{(s_1)}(\mu) \psi_{2j}^{(s_2)}(\eta) \Big ) \nonumber \\
&& \tilde{I}_{s_1,s_2}(\mu,\eta) = 2 \sum_{j=0}^{N/2 - 1} {1 \over r_j} \Big (
 \epsilon_{s_1}[\psi_{2j}^{(s_1)}](\mu) \epsilon_{s_2}[\psi_{2j+1}^{(s_2)}](\eta) -
 \epsilon_{s_1}[\psi_{2j+1}^{(s_1)}](\mu) \epsilon_{s_2}[\psi_{2j}^{(s_2)}](\eta)  \Big ) \nonumber \\
 && \qquad \qquad \qquad + \left \{ \begin{array}{ll}
 \displaystyle {\zeta \over 2} {(z^{-N_1/2} - w^{-N_1/2})^2 \over z^{-1} - w^{-1}}, &
 s_1=s_2=R, \\
 0, & {\rm otherwise}.
 \end{array} \right.
 \end{eqnarray}
 Also,  set
 \begin{equation}\label{K}
 K_{s_1,s_2}(\mu,\eta) =
 \begin{bmatrix} S_{s_1,s_2}(\mu,\eta) & - D_{s_1,s_2}(\mu,\eta) \\
 \tilde{I}_{s_1,s_2}(\mu,\eta) & S_{s_2,s_1}(\eta,\mu) \end{bmatrix}.
 \end{equation}
 (Note that $K_{GR}(y,x) = -K_{RG}(x,y)$.)
 In terms of this notation, we have from \cite{Ma10} that
 \begin{eqnarray}\label{45}
 && \rho_{(k_1,k_2)}(x_1,\dots,x_{k_1};y_1,\dots,y_{k_2})  =  {1 \over [\zeta^{N_1/2}] \prod_{l=1}^{N_1/2+N_2} r_{l-1}}
 [\zeta^{N_1/2}] \prod_{l=1}^{N_1/2+N_2} r_{l-1} \nonumber \\
 && \qquad  \times {\rm Pf} \bigg (
 \begin{bmatrix} K_{RR}(x_i,x_j) & K_{RG}(x_i,y_m) \\
 K_{GR}(y_l,x_j) & K_{GG}(y_l,y_m) \end{bmatrix}
 Z_{2(k_1+k_2)}^{-1} \bigg )_{i,j=1,\dots,k_1 \atop
 l,m=1,\dots,k_2}.
 \end{eqnarray}
 
 An important point is that the expression in the numerator of (\ref{45}) is a polynomial in
 $\zeta$. Expanding the Pfaffian for large $\zeta$ will therefore allow us to determine what powers of $\zeta$ in the expansion of $\prod_{l=1}^{N_1/2+N_2} r_{l-1}$ will combine to contribute to
 $[\zeta^{N_1/2}] $.
 
 \begin{proposition}\label{PL}
 We have
 \begin{eqnarray}\label{MR1}
&& \lim_{\zeta \to \infty} {\rm Pf} \bigg (
 \begin{bmatrix} K_{RR}(x_i,x_j) & K_{RG}(x_i,y_m) \\
 K_{GR}(y_l,x_j) & K_{GG}(y_l,y_m) \end{bmatrix}
 Z_{2(k_1+k_2)}^{-1} \bigg )_{i,j=1,\dots,k_1 \atop
 l,m=1,\dots,k_2} \nonumber \\
 && \qquad = 
{\rm Pf} \bigg (
 \begin{bmatrix} \tilde{K}_{RR}(x_i,x_j) & \tilde{K}_{RG}(x_i,y_m) \\
 \tilde{K}_{GR}(y_l,x_j) & \tilde{K}_{GG}(y_l,y_m) \end{bmatrix}
 Z_{2(k_1+k_2)}^{-1} \bigg )_{i,j=1,\dots,k_1 \atop
 l,m=1,\dots,k_2},
 \end{eqnarray}
where
\begin{equation}\label{MR1a}
\tilde{K}_{s_1,s_2}(\mu,\eta) := \lim_{\zeta \to \infty} K_{s_1,s_2}(\mu,\eta)
\end{equation}
is independent of $\zeta$.
Consequently
\begin{eqnarray}
 && \rho_{(k_1,k_2)}(x_1,\dots,x_{k_1};y_1,\dots,y_{k_2})  \nonumber \\
 && \qquad  = 
 {\rm Pf} \bigg (
 \begin{bmatrix} \tilde{K}_{RR}(x_i,x_j) & \tilde{K}_{RG}(x_i,y_m) \\
 \tilde{K}_{GR}(y_l,x_j) & \tilde{K}_{GG}(y_l,y_m) \end{bmatrix}
 Z_{2(k_1+k_2)}^{-1} \bigg )_{i,j=1,\dots,k_1 \atop
 l,m=1,\dots,k_2} \label{MR2}
 \end{eqnarray}
 \end{proposition}
 
 \noindent
 Proof. \quad Assuming (\ref{MR1}),  the Pfaffian can be expanded in inverse
 powers  of $\zeta$, with leading term a constant. But from (\ref{W3d})
 $\prod_{l=1}^{N_1/2+N_2} r_{l-1}$ is a polynomial of degree $\zeta^{N_1/2}$ and so the
 operation $[\zeta^{N_1/2}]$ in the numerator of (\ref{K}) is equal to (\ref{MR1}) times
 the denominator, and (\ref{MR2}) follows. It remains then to verify (\ref{MR1}).
 
 In fact substituting (\ref{38})--(\ref{41}) in (\ref{Ma4}) it is immediate that all the leading large
 $\zeta$ forms  are independent of $\zeta$ except for $\tilde{I}_{RR}(x,y)$.
 Noting from (\ref{W3a}), (\ref{W4c}), (\ref{38}) and (\ref{39}) that
 $$
 \epsilon_R[\psi_{2j}^{(R)}](x) =
 \left \{ \begin{array}{ll} - \pi \zeta z^{j+1-N_1}, & j=0,\dots,N_1/2-1 \\
 0, & j=N_1/2,\dots,N_1+N_2 - 1 \end{array} \right.
 $$
 \begin{equation}\label{47}
 \epsilon_R[\psi_{2j+1}^{(R)}](x) =
 \left \{ \begin{array}{ll}  \pi \zeta z^{-j}, & j=0,\dots,N_1/2-1 \\
 0, & j=N_1/2,\dots,N_1+N_2 - 1 \end{array} \right.
 \end{equation}
 we have 
 $$
 \tilde{I}_{R,R}(x,y) = -{\zeta \over 2}
 \sum_{j=0}^{N_1/2} {(2 \pi)^2 \zeta \over
 (2 \pi)^2 \zeta + 2 \pi (N_1 + 1 - 2j)} \Big (w^{j+1-N_1} z^{-j} -
 z^{j+1-N_1} w^{-j} \Big ) \: + {\zeta \over 2}
 {(w^{-N_1/2} - z^{-N_1/2})^2 \over w^{-1} - z^{-1}}.
 $$
 This would appear to be proportional to $\zeta$ for $\zeta$ large. But it is easy to verify that in fact the proportionality constant vanishes, and so
 \begin{equation}\label{MRc}
\tilde{I} _{R,R}(x,y) =   \pi \zeta 
 \sum_{j=0}^{N_1/2} {(N_1 + 1 - 2j) \over
 (2 \pi)^2 \zeta + 2 \pi (N_1 + 1 - 2j)} \Big (w^{j+1-N_1} z^{-j} -
 z^{j+1-N_1} w^{-j} \Big ).
 \end{equation}
 This like all the other entries of the Pfaffian is independent of $\zeta$ as $\zeta \to \infty$.
 \hfill $\square$

 \subsection{Bulk scaling limit}
 The probability density function (\ref{5.1}) is defined on a circle. By scaling the angles $\theta_j
 \mapsto 2 \pi x_j/L$, $\phi_\alpha \mapsto 2 \pi y_\alpha/L$ the system can instead be interpreted
 as defined on an interval of length $L$ with period boundary conditions. The thermodynamic limit is then specified as $L \to \infty$ with $N_1/L =: \rho_R$ and $N_2/L =: \rho_G$ fixed. Applied to the correlation functions it gives the bulk scaling limit, which from \cite{FJ84} has the explicit forms
 (\ref{12}) in the two-point case.
 
 Using Proposition \ref{PL} it is a straightforward exercise to compute the bulk scaling limit for the general $(k_1 + k_2)$-point correlation function. These involve the quantities
 \begin{eqnarray}\label{SID}
 && S_R^{\rm bulk}(x,y) = \rho_R \int_0^1 e^{2 \pi i \rho_R (x-y) t} \, dt \nonumber \\
 && I_R^{\rm bulk}(x,y) = \rho_R^2 \int_{-1/2}^{1/2} t e^{2 \pi i \rho_R (x-y) t} \, dt \nonumber \\
 && S_G^{\rm bulk}(x,y) = {\rho_G \over 2}
 \int_0^1 \Big ( e^{-2 \pi i (\rho_G t + \rho_R/2) (x-y) } + {\rm c.c.} \Big )\, dt \nonumber \\
 && I_G^{\rm bulk}(x,y) = {\rho_G \over 4} \int_0^1 ( \rho_G t + \rho_R/2)
 \Big ( e^{-2 \pi i (\rho_G t + \rho_R/2) (x-y) } - {\rm c.c.} \Big )\, dt \nonumber \\
  && D^{\rm bulk}(x,y) = \rho_G \int_{0}^{1} {1 \over \rho_G t + \rho_R/2}
 \Big ( e^{-2 \pi i (\rho_G t + \rho_R/2) (x-y) } - {\rm c.c.} \Big )\, dt . 
 \end{eqnarray}
 
 \begin{proposition}\label{PRG}
 Let 
 \begin{eqnarray*}
 && \rho_{(k_1,k_2)}^{\rm bulk}(x_1,\dots,x_{k_1};y_1,\dots,y_{k_2}) \nonumber \\
 && \quad = \lim_{L \to \infty}
 \Big ( {2 \pi \over L} \Big )^{k_1 + k_2}
 \rho_{(k_1,k_2)}(2 \pi x_1/L,\dots,2 \pi x_{k_1}/L;
 2 \pi y_1/L,\dots,2 \pi y_{k_2}/L) \Big |_{N_1/L = \rho_R \atop N_2/L = \rho_G}.
 \end{eqnarray*}
 We have
 \begin{eqnarray}\label{MR2bulk}
 && \rho_{(k_1,k_2)}^{\rm bulk}(x_1,\dots,x_{k_1};y_1,\dots,y_{k_2})  \nonumber \\
 && \qquad  = 
 {\rm Pf} \bigg (
 \begin{bmatrix} {K}_{RR}^{\rm bulk}(x_i,x_j) & {K}_{RG}^{\rm bulk}(x_i,y_m) \\
 {K}_{GR}^{\rm bulk}(y_l,x_j) & {K}_{GG}^{\rm bulk}(y_l,y_m) \end{bmatrix}
 Z_{2(k_1+k_2)}^{-1} \bigg )_{i,j=1,\dots,k_1 \atop
 l,m=1,\dots,k_2}
 \end{eqnarray}
 where
 \begin{equation}\label{Ka}
 K_{s_1,s_2}^{\rm bulk}(\mu,\eta) =
 \begin{bmatrix} S_{s_1,s_2}^{\rm bulk}(\mu,\eta) & - D_{s_1,s_2}^{\rm bulk}(\mu,\eta) \\
 \tilde{I}_{s_1,s_2}^{\rm bulk}(\mu,\eta) & S_{s_2,s_1}^{\rm bulk}(\eta,\mu) \end{bmatrix},
 \end{equation}
 with
 \begin{eqnarray*}
 &&S_{RR}^{\rm bulk}(x,y) = S_R^{\rm bulk}(x,y),  \qquad 
 -D_{RR}^{\rm bulk}(x,y) =-  e^{\pi i \rho_R (x+y)} D^{\rm bulk}(x,y) \\
 && \tilde{I}_{RR}^{\rm bulk}(x,y) =- e^{-\pi i \rho_R (x+y)} I_R^{\rm bulk}(x,y), \qquad S_{RG}^{\rm bulk}(x,y) =
 e^{\pi i \rho_R x} S_G(x,y) \\
 &&  -D_{RG}^{\rm bulk}(x,y) = -e^{\pi i \rho_R x} D^{\rm bulk}(x,y), \qquad 
\tilde{I}_{RG}^{\rm bulk}(x,y) = {1 \over 2} e^{-\pi i \rho_R x} I_R^{\rm bulk}(x,y) \\
&& S_{GR}(y,x) = e^{- 2 \pi i \rho_R x}  e^{  \pi i \rho_R y} S_R(x,y), \qquad
-D_{GR}^{\rm bulk}(x,y) = D_{RG}^{\rm bulk}(y,x), 
\\
&& \tilde{I}_{GR}^{\rm bulk}(x,y) = -\tilde{I}_{RG}^{\rm bulk}(y,x), \qquad
 S_{GG}^{\rm bulk}(x,y) = S_G^{\rm bulk}(x,y),  \\
&& -D_{GG}^{\rm bulk}(x,y) = -D^{\rm bulk}(x,y), \qquad
  \tilde{I}_{GG}^{\rm bulk}(x,y) = -I_G^{\rm bulk}(x,y).
 \end{eqnarray*}

 \end{proposition}
 
 \noindent
 Proof. \quad We see from (\ref{W3c}), (\ref{W3d}), (\ref{38})--(\ref{41}), together with (\ref{MR1a})
 that the summations in (\ref{Ma4}) range over $j=0,1,\dots,N_1/2$ in the cases that the summands  involve $\epsilon_R[\psi_j^{(R)}]$, and over $j=N_1/2,N_1/2+1,\dots,N_1/2+N_2-1$ otherwise.
 In the former range
 \begin{align}\label{an1}
& \psi_{2j}^{(G)}(x) = z^{-N_1/2+1+j},  &\psi_{2j+1}^{(G)}(x) = z^{N_1/2-j} \nonumber \\
&\psi_{2j}^{(R)}(x) = z^{j},  &  \psi_{2j+1}^{(R)}(x) = z^{N_1-1-j}  \nonumber  \\
&\epsilon_R[\psi_{2j}^{(R)}](x) = - \pi \zeta z^{j+1-N_1}, &
\epsilon_R[\psi_{2j+1}^{(R)}](x) =  \pi \zeta z^{-j}  \nonumber  \\
&\epsilon_G[\psi_{2j}^{(R)}](x) =  {1 \over 2} (-{N_1 \over 2} + 1 + j) z^{-N_1/2  -j}, &
\epsilon_G[\psi_{2j+1}^{(R)}](x) =   {1 \over 2} ({N_1 \over 2} - j) z^{N_1/2-j-1}  \nonumber  \\
& r_j = (2 \pi)^2 \zeta + 2 \pi (N_1+ 1 - 2j),
\end{align}
while in the latter range
 \begin{align}\label{an2}
& \psi_{2j}^{(R)}(x) = z^{-N_2+j-N_1/2}, & \psi_{2j+1}^{(R)}(x) = z^{N_2+3N_1/2-j-1} \nonumber   \\
&\psi_{2j}^{(G)}(x) = z^{j-N_1-N_2+1},  & \psi_{2j+1}^{(G)}(x) = z^{N_1 + N_2 -j}  \nonumber \\
&\epsilon_G[\psi_{2j}^{(G)}](x) = {1 \over 2} (j - N_1 - N_2) z^{j-N_1-N_2}, 
&\epsilon_G[\psi_{2j+1}^{(G)}](x) =  {1 \over 2} (N_1 + N_2-j) z^{N_1+N_2-j-1}  \nonumber \\
& r_j = 2 \pi (2N_2+ 2N_1-1 - 2j),
\end{align}

In relation to the cases that the summand contains $\epsilon_R[\psi_j^{(R)}]$, appropriate use of
(\ref{an1}) substituted in the first and third formulas of (\ref{Ma4}), or (\ref{MRc}), shows that the summations are in fact Riemann sums. Simple manipulations then give
\begin{eqnarray}
&& \lim_{L \to \infty} {2 \pi \over L} S_{RR}(2 \pi x/L, 2 \pi y/L) =
\rho_R \int_0^1 e^{2 \pi i \rho_R (x - y) t} \, dt \nonumber \\
&& \lim_{L \to \infty} {2 \pi \over L} S_{GR}(2 \pi x/L, 2 \pi y/L) =
\rho_R e^{- \pi i \rho_R y} \int_0^1 e^{-2 \pi i \rho_R (x - y) t} \, dt \nonumber \\
&& \lim_{L \to \infty} {2 \pi \over L^2} \tilde{I}_{RG}(2 \pi x/L, 2 \pi y/L) =
{1 \over 2} \rho_R^2 e^{- \pi i \rho_R x} \int_{-1/2}^{1/2} t e^{2 \pi i \rho_R (x - y) t} \, dt 
 \nonumber \\
&& \lim_{L \to \infty} {2 \pi \over L^2} \tilde{I}_{RR}(2 \pi x/L, 2 \pi y/L) = -
 \rho_R^2 e^{- \pi i \rho_R (x+y)} \int_{-1/2}^{1/2} t e^{2 \pi i \rho_R (x - y) t} \, dt 
\end{eqnarray}
Similarly, in the cases that the summand does not contain $\epsilon_R[\psi_j^{(R)}]$ we find
\begin{eqnarray}\label{54}
&& \lim_{L \to \infty} {2 \pi \over L} S_{GG}(2 \pi x/L, 2 \pi y/L) = 
{\rho_G \over 2} \int_0^1 \Big ( e^{-2 \pi i (\rho_G t + \rho_R/2)(x-y)} + {\rm c.c.} \Big ) \, dt
\nonumber \\
&& \lim_{L \to \infty} {2 \pi \over L} S_{RG}(2 \pi x/L, 2 \pi y/L) = 
{\rho_G \over 2} e^{\pi i \rho_R x} \int_0^1 \Big ( e^{-2 \pi i (\rho_G t + \rho_R/2)(x-y)} + {\rm c.c.} \Big ) \, dt
\nonumber \\
&&\lim_{L \to \infty} {2 \pi \over L^2} \tilde{I}_{GG}(2 \pi x/L, 2 \pi y/L) = 
-{\rho_G \over 4} \int_0^1 (\rho_G t + \rho_R/2) \Big ( e^{-2 \pi i (\rho_G t + \rho_R/2)(x-y)} - {\rm c.c.} \Big ) \, dt
\nonumber \\
&& \lim_{L \to \infty} 2 \pi D_{GG}(2\pi x/L, 2 \pi y/L) = \rho_G  \int_0^1 {1 \over
\rho_G t + \rho_R/2} \Big ( e^{-2 \pi i (\rho_G t + \rho_R/2)(x-y)} - {\rm c.c.} \Big ) \, dt \nonumber \\
&& \lim_{L \to \infty} 2 \pi D_{RG}(2\pi x/L, 2 \pi y/L) = 
\rho_G e^{ \pi i \rho_R x} \int_0^1 {1 \over \rho_G t  + \rho_R/2}
\Big ( e^{-2 \pi i (\rho_G t + \rho_R/2) (x-y)} - {\rm c.c.} \Big ) \, dt \nonumber \\
&& \lim_{L \to \infty} 2 \pi D_{RR}(2\pi x/L, 2 \pi y/L) = 
\rho_G e^{ \pi i \rho_R (x+y)} \int_0^1 {1 \over  \rho_G t + \rho_R/2}
\Big ( e^{-2 \pi i (\rho_G t + \rho_R/2) (x-y)} - {\rm c.c.} \Big ) \, dt  \nonumber \\
\end{eqnarray}
Upon recalling the definitions (\ref{SID}), and making use of the fact that the Pfaffian and thus
correlation function is unchanged if the element $D_{s_1,s_2}$ in (\ref{K}) is multiplied by
the scalar $L$ say, and $\tilde{I}_{s_1,s_2}$ is multiplied by  $1/L$ before the limit
$L \to \infty$ is taken, we see that the stated forms follow. \hfill $\square$

\section{Properties of the correlations}
\subsection{Explicit form of the two-point functions}
Denoting the bulk two-point function by $\rho_{s_1 s_2}(x,y)$ we see from (\ref{MR2bulk}) and (\ref{Ka}) that
$$
\rho_{s_1 s_2}(x,y) = \rho_{s_1} \rho_{s_2} - \Big ( S_{s_1,s_2}(x,y) S_{s_2,s_1}(y,x) +
D_{s_1,s_2}(x,y) \tilde{I}_{s_1,s_2}(x,y) \Big ).
$$
Substituting the explicit forms of the quantities on the right hand side from Proposition
\ref{PRG} we can check, after some minor manipulation, that there is agreement with the forms
(\ref{12}). Furthermore, in the case $\rho_G=0$, we have from (\ref{SID}) that 
$I_R^{\rm bulk}(x,y) = 0$. The structure of (\ref{MR2bulk}) in the case $k_2=0$, and in particular the fact already alluded to below (\ref{54}) that in the expansion of the Pfaffian each
factor of $\tilde{I}^{\rm bulk}$ is multiplied by $D^{\rm bulk}$ tells us that we can effectively  set
$D^{\rm bulk}(x,y)=0$. The Pfaffian then reduces to a determinant,
 and this upon recalling the explicit form of
$S_R^{\rm bulk}(x,y)$ from (\ref{SID}) is seen to be precisely the  determinant formula (\ref{3}) 
for $\rho^{\rm bulk}_{(k,0)}$. 

We can check too that with $\rho_R=0$ the Pfaffian formula  (\ref{4}) for
$\rho^{\rm bulk}_{(0,k)}$ is reclaimed.
\subsection{Sum rules}
Let us now turn our attention to the screening sum rules (\ref{13}). In physical terms these say that upon fixing a Roman species particle, the other Roman species particles collectively redistribute to exactly cancel out the density of this particle. Similarly for the Greek species. This behaviour is to be contrasted to what happens in the two-component plasma specified by (\ref{4.2}). There the analogue of (\ref{13}) reads
\begin{equation}\label{13a}
\int_{-\infty}^\infty \Big ( \rho_{+1,+1}^T(x,0) + 2 \rho_{+1,+2}^T(x,0) \Big ) \, dx = - \rho_{+1},
\quad
\int_{-\infty}^\infty \Big ( \rho_{+2,+1}^T(x,0) + 2 \rho_{+2,+2}^T(x,0) \Big ) \, dx = -2 \rho_{+2}.
\end{equation}
In particular, unlike with (\ref{13}), there is no special form for the individual integrals.
In words the sum rules (\ref{13a}) say that the effect of fixing a single $+1$ charge species is for both the $+1$ and $+2$ charges to collectively respond to exactly cancel out this charge density, and similarly the effect of fixing a $+2$ charge species.

Notice that (\ref{13}) and (\ref{13a}) involve the truncated two particle correlations.
In general 
the $(k_1,k_2)$-point correlation $\rho^{\rm bulk}_{(k_1,k_2)}$ does not decay for large distances in any of its arguments and consequently cannot be integrated over $\mathbb R$ with respect to any one of its arguments. However, by adding suitable linear combinations 
of lower order correlations to $\rho_{(k_1,k_2)}^{\rm bulk}$ a quantity 
$\rho_{(k_1,k_2)}^{\rm bulk T}$ --- the fully truncated $(k_1,k_2)$-point correlation -- can be obtained, which has the property of decaying when any one particle coordinate is taken to infinity. The simplest case is when $k_1 + k_2 = 2$ and we have for example
$\rho_{(1,1)}^T(x,0) = \rho_{(1,1)}(x,0) - \rho_{(1,0)}
\rho_{(0,1)}$. A significant feature of the Pfaffian form (\ref{MR2})
for the correlation functions is that it allows for
$\rho_{(k_1,k_2)}^{\rm bulk \, T}$
in the general case to be written in the structured form \cite[Prop.~5.1.2]{Fo10}
\begin{eqnarray}\label{14}
&& \rho_{(k_1,k_2)}^{\rm bulk \, T}(x_1,\dots,x_{k_1}; y_1,\dots,y_{k_2})= (-1)^{k_1+k_2-1} 
\sum_{{\rm cycles} \atop {\rm length} \: k_1 + k_2}\nonumber \\
&& \quad  \times
\Big ( K_{s(z_{i_1}), s(z_{i_2})}(y_{i_1}, y_{i_2}) K_{s(z_{i_2}), s(z_{i_3})}(y_{i_2}, y_{i_3}) 
\cdots K_{s(z_{i_{k_1+ i_{k_2})}, s(z_{i_1})}}(y_{i_{k_1+ i_{k_2}}}, y_{i_1}) \Big )^{(0)}
\end{eqnarray}
where $\{z_i\}_{i=1,\dots,k_1 + k_2} = \{x_i \}_{i=1,\dots,k_1} \cup
 \{y_j \}_{j=1,\dots,k_1}$, $s(z)$ refers to the species of coordinate $z$ (thus $s(x) = R$,
 $s(y) = G$) and the operator $( \: \: )^{(0)}$ refers to ${1 \over 2} {\rm Tr}$.
 
 For the two-component plasma the general truncated $(k_1+k_2)$-point correlation satisfies a generalization of the sum rules (\ref{13a}) \cite[(14.20)]{Fo10},
 \begin{eqnarray}\label{13b}
 && \int_{-\infty}^\infty \rho_{(k_1+1,k_2)}^T(x_1,\dots,x_{k_1},x;
 y_1,\dots,y_{k_2}) \, dx \nonumber \\
 && \qquad \qquad + 2  \int_{-\infty}^\infty \rho_{(k_1,k_2+1)}^T(x_1,\dots,x_{k_1};
 y_1,\dots,y_{k_2},y) \, dy  \nonumber \\
 && \quad = - (k_1 + 2 k_2)  \rho_{(k_1+1,k_2)}^T(x_1,\dots,x_{k_1},x;
 y_1,\dots,y_{k_2}).
 \end{eqnarray}
 Here the factors of 2 are due to the charge density being the fundamental quantity.
 This sum rule is equivalent to the requirement that upon fixing $+1$ charges at
 $x_1,\dots,x_{k_1}$ and $+2$ charges at $ y_1,\dots,y_{k_2}$, the charge density of the system responds to exactly cancel these fixed charges.
 
 We know that for the generalized plasma, in the case of the two-point functions,  the stronger sum rules (\ref{13}) hold true.  We will see that the existence of stronger screening rules is restricted
 to $k_1=0$, while  more generally
 \begin{eqnarray}\label{13c}
 && \int_{-\infty}^\infty \rho_{(k_1+1,k_2)}^T(x_1,\dots,x_{k_1},x;
 y_1,\dots,y_{k_2}) \, dx \nonumber \\
 && \qquad \qquad +   \int_{-\infty}^\infty \rho_{(k_1,k_2+1)}^T(x_1,\dots,x_{k_1};
 y_1,\dots,y_{k_2},y) \, dy  \nonumber \\
 && \quad = - (k_1 +  k_2)  \rho_{(k_1+1,k_2)}^T(x_1,\dots,x_{k_1},x;
 y_1,\dots,y_{k_2}).
 \end{eqnarray}
 In contrast to (\ref{13b}) there are no factors of 2 indicating that screening is due to the total particle density rather than total charge density.
 The sum rule (\ref{13c}) is corollary of the following integration formulas for the matrix kernels (cf.~\cite[Prop.~4.7]{FM09}).
 
 \begin{proposition}
 We have
 \begin{eqnarray}\label{54a}
 \int_{-\infty}^\infty \Big (K_{RR}(x_1,u) K_{RR}(u,x_2) +  K_{RG}(x_1,u) K_{GR}(u,x_2) \Big )\, du=
 K_{RR}(x_1,x_2) \nonumber  \\
  \int_{-\infty}^\infty   K_{GG}(y_1,u) K_{GG}(u,y_2) \, du=
 K_{GG}(y_1,y_2) \nonumber \\
   \int_{-\infty}^\infty K_{GR}(y_1,u) K_{RG}(u,y_2) \, du = 0 \nonumber \\
  \int_{-\infty}^\infty \Big (K_{GR}(y,u) K_{RR}(u,x) +  K_{GG}(y,u) K_{GR}(u,x) \Big )\, du=
  K_{GR}(y,x) \nonumber \\
  \int_{-\infty}^\infty \Big (K_{RR}(x,u) K_{RG}(u,y) +  K_{RG}(x,u) K_{GG}(u,y) \Big ) \, du=
 K_{RG}(x,y).
 \end{eqnarray} 
  \end{proposition}
 
 \noindent
 Proof. \quad The main formula required in the derivations is the generalized integral
 $$
 \int_{-\infty}^\infty e^{2 \pi i k s} \, dk = \delta(s)
 $$
 where $\delta(s)$ is the Dirac delta function. For example, using this formula we obtain
  \begin{equation}
 \int_{-\infty}^\infty K_{RR}(x,y) K_{RR}(y,x') \, dy =
 \begin{bmatrix} S_R(x,x') & 0 \label{F1} \\
 - e^{-\pi i \rho_R(x+x')} I_R(x,x') & S_R(x',x) \end{bmatrix} 
 \end{equation}
 and
 \begin{equation}\label{F2}
 \int_{-\infty}^\infty K_{RG}(x,y) K_{GR}(y,x') \, dy =
 \begin{bmatrix} 0 & - {1 \over 2} e^{\pi i \rho_R(x+x')} D(x,x') \\
 0 & 0 \end{bmatrix}.
  \end{equation}
  Adding (\ref{F1}) to  (\ref{F2}) and recalling (\ref{54}) we obtain the first of the sum rules
  in (\ref{54a}).
  
  The derivation of the remaining formulas in (\ref{54a}) proceeds similarly. \hfill $\square$
  
  \smallskip
  We see that indeed that after substituting (\ref{14}) in the LHS of
  (\ref{13b}) and using the integration formulas (\ref{54a}) that the RHS of (\ref{13c}) results. In the
  case of species $G$ only, 
  we similarly deduce the stronger result
   \begin{equation}
    \int_{-\infty}^\infty \rho_{(0,k_2+1)}^T(
 y_1,\dots,y_{k_2},y) \, dy
  = -  k_2  \rho_{(0,k_2)}^T(
 y_1,\dots,y_{k_2}).
 \end{equation}
  On the other hand, in the case of species $R$ only, while
  $\rho_{RR}^T$  satisfies the stronger
  sum rule in  (\ref{13}), it follows from (\ref{F1}) that we do not have for example that
  $$
  \int_{-\infty}^\infty \rho_{RRR}^T(x_1,x_2,x) \, dx
  $$
  is equal to $-\rho_{RR}^T(x_1,x_2)$. 
  
  \subsection{The large density limit of one of the species}
  A possible limit of the two-component system is to take the density of one of the species to infinity, while keeping the density of the other species fixed. It turns out that this limit applied to the two-point correlations reveals a dramatic difference between the true log-potential plasma system (\ref{4.2}) and the generalized plasma
  (\ref{5.1}). 
  
  For the generalized plasma, we see from (\ref{10}) and (\ref{11}) that
 \begin{eqnarray*}
&&
\lim_{\rho_G \to \infty} \rho_{RR}^T(x,0) =  - {\sin^2 \pi \rho_R x \over (\pi x)^2}  \nonumber \\
&& \qquad \qquad -
 \rho_R^2 \int_0^\infty  dt \int_0^1 ds \, {1 - 2s \over 2  t + 1}
\Big ( e^{2 \pi i \rho_R x (s - 1) - 2 \pi i \rho_R x t} -
e^{2 \pi i \rho_R x s + 2 \pi i \rho_R xt} \Big ) \label{10a}\\
&&\lim_{\rho_R \to \infty}  \rho_{GG}^T(x,0) = -  {\sin^2 \pi \rho_G x \over (\pi x)^2}.
\end{eqnarray*}
On the other hand, the explicit form of the two-point correlations for the 
true log-potential plasma system (\ref{4.2}) \cite{Fo84a} allows us to compute that 
$$
\lim_{\rho_{+2} \to \infty} \rho_{+1,+1}^T(x,0) = 0, \qquad
\lim_{\rho_{+1} \to \infty} \rho_{+2,+2}^T(x,0) = 0,
$$
where it is assumed $x \ne 0$. Thus in the latter case the fixed density species become uncorrelated when placed in a sea of the other species. This contrasts to the behaviour exhibited by  the generalized plasma, in which for this setting the fixed density species still exhibits slowly decaying correlations.

Some insight into the reason for this is that, in the notation for the couplings of the generalized plasma introduced below (\ref{5.1}), arguments based on the direct correlation function were used to obtain the predictions for the large distance asymptotic forms
$$
\rho_{RR}^T(x,0) \sim - {g_{RR} \over \pi^2 \Delta x^2}, \qquad \rho_{RG}^T(x,0) \sim
{g_{RG} \over \pi^2 \Delta x^2}, \qquad
\rho_{GG}^T(x,0) \sim - {g_{GG} \over \pi^2 \Delta x^2}
$$
where $\Delta := g_{RR} g_{GG} - g_{RG}^2$. These are independent of the particle density. In contrast, for the 
true log-potential plasma system, charges $+1$ and $+2$, it is only the combination
$$
\rho_{+1,+1}^T(x,0) + 4 \rho_{+1,+2}^T(x,0) + 4 \rho_{+2,+2}^T(x,0) \sim - {1 \over \pi^2 x^2}
$$
relating to the charge-charge correlation which exhibits a density independent large $x$ form.

\subsection*{Acknowledgements}
 This work was done while the authors were members of the MSRI,
 participating in the Fall 2010 program `Random matrices, interacting
 particle systems and integrable systems'. Partial support for the
 research of PJF was also provided by the Australian Research
 Council. The research of CDS was partially supported by the (United
 States) National Science Foundation (DMS-0801243).   

\appendix
 
 \section{Appendix}

In \cite{IO06}, Ishikawa, Okada, Tagawa and Zeng give the
following Pfaffian identity for the Vandermonde determinant.
\begin{theorem}
\label{thm:1}
Suppose $N$ is an even integer and let $\mathbf x =(x_1, x_2, \ldots,
x_N)$ be indeterminants.  Then,
\begin{equation}
\label{eq:1}
\Pf \left[ \frac{\big(x_n^{N/2} - x_m^{N/2}\big)^2}{(x_n - x_m)}
\right]_{m,n=1}^N = \prod_{1 \leq m < n \leq N} (x_n - x_m).
\end{equation}
\end{theorem}
Ishikawa {\it et al.}~prove this identity as a special case of a more
general Pfaffian evaluation.  A simpler proof, following that of the
Vandermonde determinant identity, is outlined as follows:
\begin{enumerate}
\item The left hand side of (\ref{eq:1}) is 0 if $x_n = x_m$ for any
  $n \neq m$, and thus the right hand side divides the left (as a
  polynomial in $\Q[x_1, x_2, \ldots, x_N]$;
\item The polynomials defined by the left and right hands sides of
  (\ref{eq:1}) are homogeneous and of the same degree, and thus their
  ratio is a non-zero rational number;
\item This rational number is equal to 1 as seen by checking that the 
  coefficients on the left and right hand side of any particular
  monomial are equal.  
\end{enumerate}
This proof works more generally.  If $F(x)$ and $G(x)$ are  monic
polynomials of degree $N/2$, then 
\[
\Pf \left[ \frac{\big(F(x_n) - F(x_m)\big)\big(G(x_n) - G(x_m)\big)}{(x_n - x_m)}
\right]_{m,n=1}^N = \prod_{1 \leq m < n \leq N} (x_n - x_m).
\] 

The following Theorem is similar in spirit to these Pfaffian
evaluations, but is in a form which lends itself more naturally to
calculations which arise in random matrix theory and related fields. 
\begin{theorem}
\label{thm:2}
Suppose $N$ is an even integer and let $\mathbf x =(x_1, x_2, \ldots,
x_N)$ be indeterminants.
If $\pi_0, \pi_1, \ldots, \pi_{N-1}$ are polynomials with
  $\deg \pi_n = n$ and leading coefficients $a_0, a_1, \ldots, a_{N-1}$, then
\[
\Pf \left[ 
\sum_{\ell=0}^{N/2-1} \pi_{2\ell}(x_m) \pi_{2\ell+1}(x_n) -
\pi_{2\ell+1}(x_m) \pi_{2\ell}(x_n)
\right]_{m,n=1}^N = \prod_{k=0}^{N-1} a_k \prod_{1 \leq m < n \leq N} (x_n - x_m). 
\]
\end{theorem}

In addition, we prove a confluent form of this identity. For each
non-negative integer $\ell$ we define the differential operator 

\[
D_{\ell} f(x) = \frac{1}{(\ell-1)!}
\frac{d^{\ell}}{dx^{\ell}} f(x).
\] 
($D_0$ is the identity operator).  If $L$ is a positive integer and
$\pi_0, \pi_1, \ldots, \pi_{LN-1}$ are polynomials with $\deg p_n =
n$ and leading coefficients $a_0, a_1, \ldots, a_{LN-1}$,   we define 
the $LN \times LN$ confluent Vandermonde matrix 
\begin{equation}
\label{eq:3}
V = \begin{bmatrix}
D_0 \pi_{n-1}(x_m) \\
D_1 \pi_{n-1}(x_m) \\ 
\vdots \\
D_{L-1} \pi_{n-1}(x_m) \\
\end{bmatrix}; \qquad {m = 1,2,\ldots, N; \atop n = 1, 2, \ldots LN.}
\end{equation}
The confluent Vandermonde identity has that
\[
\det V = \prod_{k=0}^{LN-1} a_k \prod_{1 \leq m < n \leq N} (x_n - x_m)^{L^2}.
\]

\begin{theorem}
\label{thm:3}
Suppose $L$ is an integer, $N$ is an even integer and $\pi_0, \pi_1,
\ldots, \pi_{LN-1}$  are 
polynomials with $\deg p_n = n$ and leading coefficients $a_0, a_1,
\ldots, a_{LN-1}$.  Define,
\begin{equation}
\label{eq:4}
K(x,y) = \sum_{n=0}^{LN/2-1} \pi_{2n}(x) \pi_{2n+1}(y) - \pi_{2n+1}(x)
\pi_{2n}(y). 
\end{equation}
and for each $0 \leq \ell, m \leq L-1$ define 
\[
D^{\ell} K(x,y) D^m = \sum_{n=0}^{LN/2-1} D^{\ell} \pi_{2n}(x) D^m
\pi_{2n+1}(y) - D^{\ell} \pi_{2n+1}(x) D^m \pi_{2n}(y).
\]
Define the antisymmetric $LN \times LN$ matrix 
\[
W = \begin{bmatrix}
D^0 K(x_m,x_n) D^0 & D^0 K(x_m,x_n) D^1 &  & D^0 K(x_m,x_n)
D^{L-1} \\
D^1 K(x_m,x_n) D^0 & D^1 K(x_m,x_n) D^1 & \cdots & D^1 K(x_m,x_n)
D^{L-1} \\
& \vdots & \ddots & \vdots \\
D^{L-1} K(x_m,x_n) D^0 & D^{L-1} K(x_m,x_n) D^1 & \cdots & D^{L-1} K(x_m,x_n)
D^{L-1} 
\end{bmatrix}_{m,n=1}^N.
\]
Then,
\[
\Pf W = \prod_{k=0}^{LN-1} a_k \prod_{1 \leq m < n \leq N} (x_n - x_m)^{L^2}.
\]
\end{theorem}

Clearly Theorem~\ref{thm:2} follows from Theorem~\ref{thm:3} by
setting $L = 1$.  

\begin{cor}
\label{cor:1}
Let
\[
F(x,y) = \frac{(y^{LN/2} - x^{LN/2})^2}{(y-x)},
\]
and for each $0 \leq \ell, m \leq L-1$ let 
\[
D^{\ell} F(x,y) D^m = \frac{1}{(\ell-1)! (m-1)!} \left( \lim_{w
    \rightarrow x} \lim_{z \rightarrow y}  \frac{\partial^{\ell + m}}{\partial
  w^{\ell} \partial z^m} F(w,z) \right),
\]
and define the $LN \times LN$ antisymmetric matrix 
\[
U = \begin{bmatrix}
D^0 F(x_m,x_n) D^0 & D^0 F(x_m,x_n) D^1 &  & D^0 F(x_m,x_n)
D^{L-1} \\
D^1 F(x_m,x_n) D^0 & D^1 F(x_m,x_n) D^1 & \cdots & D^1 F(x_m,x_n)
D^{L-1} \\
& \vdots & \ddots & \vdots \\
D^{L-1} F(x_m,x_n) D^0 & D^{L-1} F(x_m,x_n) D^1 & \cdots & D^{L-1} F(x_m,x_n)
D^{L-1} 
\end{bmatrix}_{m,n=1}^N.
\]
Then,
\[
\Pf U = \prod_{1 \leq m < n  \leq N} (x_n - x_m)^{L^2}. 
\]
\end{cor}

\noindent Proof of Theorem~\ref{thm:3}. \quad We define the $LN \times
LN$ matrix 
\[
J = \begin{bmatrix}
0 & 1 \\
-1 & 0 \\
& & 0 & 1 \\
& & -1 & 0 \\
& & & & \ddots \\
& & & & & 0 & 1 \\
& & & & & -1 & 0 \\
\end{bmatrix}.
\]
It is an easy exercise to check that $\Pf J = 1$.  A similarly easy
calculation shows that $W = V J V^{\transpose}$, and thus, using
properties of the Pfaffian, $\Pf W = \det V \Pf J = \det V.$ \hfill
$\square$

\noindent Proof of Corollary~\ref{cor:1}. \quad Suppose we permute and
relabel $\pi_0, \pi_1, \ldots, \pi_{LN-1}$, so 
that it is not necessarily the case the $\deg \pi_n = n$.  If $\sigma$
is the permutation of $0,1,\ldots, LN-1$ which we used to permute our
polynomials, then it is easy to see that $V$, formed as in
(\ref{eq:3}), but using our reordered and relabeled polynomials will
satisfy 
\[
\det V = \sgn \sigma \prod_{k=0}^{LN-1} a_k \prod_{1 \leq m < n \leq
  N} (x_n - x_m)^{L^2},
\]
and if we define $K(x,y)$ as in (\ref{eq:4}), but with our reordered
and relabeled polynomials and defining $W$ with this `new' $K$,
we have 
\[
\Pf W  = \sgn \sigma \prod_{k=0}^{LN-1} a_k \prod_{1 \leq m < n \leq
  N} (x_n - x_m)^{L^2}.
\]
Now, 
\[
F(x,y) = \frac{(y^{LN/2} - x^{LN/2})^2}{(y-x)} = \sum_{n=0}^{LN/2-1}
y^{LN-1-n} x^n - y^{n} x^{LN-1-n}.
\]
Thus, if we define
\[
\pi_{2n+1}(x) = x^{LN-1-n} \qquad \mbox{and} \qquad \pi_{2n}(x) = x^n; \qquad
n=0,1,\ldots, LN/2-1,
\]
we have 
\[
F(x,y) = \sum_{n=0}^{LN/2-1}  \pi_{2n}(x) \pi_{2n+1}(y) -
\pi_{2n+1}(x) \pi_{2n}(y) = K(x,y).
\]
Finally, the signature of the permutation which orders $\pi_0, \pi_1,
\ldots, \pi_{LN-1}$ by degree can be shown to always be 1 (since the
reverse permutation on $LN/2$ elements has the same signature as the
perfect shuffle on $LN$ elements).  \hfill $\square$

%\bibliographystyle{amsplain}
%\bibliography{book1}

\input{ForSin22Nov.bbl}

\end{document}

%% file: ForSin22Nov.bbl
\providecommand{\bysame}{\leavevmode\hbox to3em{\hrulefill}\thinspace}
\providecommand{\MR}{\relax\ifhmode\unskip\space\fi MR }
\providecommand{\MRhref}[2]{%
  \href{http://www.ams.org/mathscinet-getitem?mr=#1}{#2}
}
\providecommand{\href}[2]{#2}